\documentclass[11pt,english]{article}
\usepackage{amssymb, amsmath}
\usepackage{multicol}
\usepackage{amsmath, amssymb}
\usepackage[mathscr]{eucal}
\usepackage{graphicx}
\usepackage{amsfonts}
\usepackage{babel}
\newcommand{\bfi}{\bfseries\itshape}
\newcommand{\rem}[1]{}
\newcommand{\remfigure}[1]{}
\makeatletter
\@addtoreset{equation}{section}

\makeatother

\def\contract{\makebox[1.2em][c]{\mbox{\rule{.6em}
{.01truein}\rule{.01truein}{.6em}}}}

\newtheorem{theorem}{Theorem}

\numberwithin{theorem}{section}

\def\0{{\bf 0}}

\begin{document}

\title{
%
%Singular solutions for geodesic flows of Vlasov moments
Vlasov moments, integrable systems and singular solutions \\
%with applications to fluids and accelerator beams
}
\author{\vspace{2mm}
J. Gibbons$^{1}$, D. D. Holm$^{1,\,2}$ and C. Tronci$^{1,3}\!$\\
%\footnote{
%$\,$also at {\it TERA Foundation for Oncological Hadrontherapy, 11 Via Puccini, %Novara 28100, Italy}
%}
\\
{\small $^1$ \it Department of Mathematics, Imperial College London, London SW7 2AZ, UK}\\
{\small $^2$ \it Computer and Computational Science Division, Los Alamos National Laboratory,} \\{\small\it Los Alamos, NM, 87545 USA} \\
{\small $^3$\,\it TERA Foundation for Oncological Hadrontherapy,
11 V. Puccini, Novara 28100, Italy}
\\ \\
}
%\date{For Henry McKean, on the occasion of his 75th birthday\\}
\date{}

\maketitle

\begin{abstract}\noindent
The Vlasov equation for the collisionless evolution of the single-particle
probability distribution function (PDF) is a well-known Lie-Poisson Hamiltonian system. Remarkably, the operation of taking the moments of the Vlasov PDF preserves the Lie-Poisson structure. The individual particle motions correspond to singular solutions of the Vlasov equation. The paper focuses on \emph{singular} solutions of the problem of {\it geodesic} motion of the Vlasov moments. These singular solutions recover geodesic motion of the individual particles. 
\rem{
The Vlasov equation for the collisionless evolution of the single-particle
probability distribution function (PDF) is a well-known example of
coadjoint motion. Remarkably, the property of coadjoint motion survives
the process of taking moments. That is, the evolution of the moments of
the Vlasov PDF is also coadjoint motion. We find that {\it geodesic}
coadjoint motion of the Vlasov moments with respect to powers of the
single-particle momentum admits singular (weak) solutions concentrated on
embedded subspaces of physical space. The motion and interactions of these
embedded subspaces are governed by canonical Hamiltonian equations for
their geodesic evolution. 
By symmetry under exchange of canonical
momentum and position, the Vlasov moments with respect to powers of the
single-particle position admit singular (weak) solutions concentrated on
embedded subspaces of momentum space.}\\
\end{abstract}

\tableofcontents

%%%%%%%%%%%%%%%%
\section{Introduction}

\paragraph{The Vlasov equation.}

The evolution of $N$ identical particles in phase space with coordinates 
$(q_i, p_i)$ $i=1,2,\dots,N$, may be described by an evolution equation
for their joint probability distribution function. Integrating over all
but one of the particle phase-space coordinates yields an evolution
equation for the single-particle probability distribution function (PDF).
This is the Vlasov equation, which may be expressed as an
advection equation for the phase-space density $f$ along the Hamiltonian vector field ${\bf X}_H$  corresponding to single-particle motion with Hamiltonian $H(q,p)$:
\begin{equation}\label{advect}
\frac{\partial f}{\partial t} 
\,=
\big\{f\,,\,H\big\}
=
\,- \,\text{\large div}_{(q,p)}(f\, {\bf X}_H)\,
= - \,\mathcal{L}_{\mathbf{X}_{\!H}}\, f
\qquad\text{ with }\quad
%$\mathcal{L}$ stands for Lie derivative and
{\bf X}_H (q,p)= \left(\frac{\partial H}{\partial p},-\frac{\partial H}{\partial q}\right)
%\qquad \textnormal{with }\quad
%h(q,p)=\frac{p^2}{2}+\phi(q)
\end{equation}

The solutions of the Vlasov equation reflect its heritage in particle
dynamics, which may be reclaimed by writing its many-particle PDF as a
product of delta functions in phase space. Any number of these delta
functions may be integrated out until all that remains is the dynamics of
a single particle in the collective field of the others. 
\rem{
In plasma
physics, this collective field generates the total electromagnetic
properties and the self-consistent equations obeyed by the single
particle PDF are the Vlasov-Maxwell equations. 
}

In the mean-field approximation of plasma dynamics, this collective field generates the total electromagnetic properties and the self-consistent equations obeyed by the single particle PDF are the Vlasov-Maxwell equations. In the electrostatic approximation, these reduce to the Vlasov-Poisson (VP) equations, which govern the statistical distributions of particle systems ranging from integrated circuits (MOSFETS, metal-oxide semiconductor field-effect transistors), to charged-particle beams, to the distribution of stars in a galaxy. 

A class of singular solutions of the VP equations called the ``cold
plasma'' solutions have a particularly beautiful experimental realization
in the Malmberg-Penning trap. In this experiment, the time average of
the vertical motion closely parallels the Euler fluid equations. In
fact, the cold plasma singular Vlasov-Poisson solution turns out to obey
the equations of point-vortex dynamics in an incompressible ideal flow.
This coincidence allows the discrete arrays of ``vortex crystals''
envisioned by J. J. Thomson for fluid vortices to be realized
experimentally as solutions of the Vlasov-Poisson equations. For a
survey of these experimental cold-plasma results see \cite{DuON1999}. 

\paragraph{Vlasov moments.}

The Euler fluid equations arise by imposing a closure relation
on the first three momentum moments, or $p-$moments of the Vlasov PDF
$f(p,q,t)$. The zero-th $p-$moment is the spatial density of particles.
The first $p-$moment is the mean fluid momentum.  Introducing an expression for the fluid pressure in terms of the density and momentum closes the system of $p-$moment equations, which otherwise would possess a countably infinite number of dependent variables. 

\rem{
The operation of taking $p-$moments preserves the geometric nature of
Vlasov's equation. In particular, this operation is a Poisson map that takes
the Vlasov Lie-Poisson structure into another system of Lie-Poisson type,
although it is not simple coadjoint motion, as for the Vlasov
PDF.}

The operation of taking $p-$moments preserves the geometric nature of Vlasov dynamics. In particular, this operation is a Poisson map. That is, it takes the Lie-Poisson structure for Vlasov dynamics into another Lie-Poisson system. However, strictly speaking, the solutions for the $p-$moments cannot yet be claimed to undergo coadjoint motion, as in the case of the Vlasov PDF solutions, because the group action underlying the Lie-Poisson structure found for the $p-$moments is not yet understood.

A closure after the first $p-$moment results in
Euler's useful and beautiful theory of ideal fluids which is also Lie-Poisson. As its primary geometric
characteristic, Euler's fluid theory represents fluid flow as
Hamiltonian geodesic motion on the space of smooth invertible maps
acting on the flow domain and possessing smooth inverses. The (left) action
of these smooth
maps (called diffeomorphisms) on the fluid reference configuration  moves the fluid particles around in their container. And their smooth
inverses recall the initial reference configuration (or label) for the
fluid particle currently occupying any given position in space. Thus, the
motion of all the fluid particles in a container is represented as a
time-dependent curve in the infinite-dimensional group of
diffeomorphisms. Moreover, this curve describing the sequential actions
of the diffeomorphisms on the fluid domain is a special optimal curve that
distills the fluid motion into a single statement. Namely, ``A fluid moves
to get out of its own way as efficiently as possible''. Put more
mathematically, fluid flow occurs along a curve in the diffeomorphism
group which is a geodesic with respect to the metric on its tangent space
supplied by its kinetic energy. 

Given the beauty and utility of the solution behavior for Euler's
equation for the first $p-$moment, one is intrigued to know more about
the dynamics of the other moments of Vlasov's equation. Of course, the
dynamics of the the $p-$moments of the Vlasov-Poisson equation is one of  the
mainstream subjects of plasma physics and space physics.

\paragraph{Summary.}
This paper formulates the dynamics of Vlasov $p-$moments governed by {\it quadratic}
Hamiltonians. This dynamics is a certain type of geodesic motion on the
symplectomorphisms, rather than on the diffeomorphisms for fluids. The
symplectomorphisms are smooth invertible maps acting on the phase
space and possessing smooth inverses. 
\rem{
The theory of moment dynamics for the Vlasov equation turns out to be connected to integrable shallow water equations, in particular, to the Benney long wave equation \cite{Be1973, Gi1981}.
}
The theory of moment dynamics
for the Vlasov equation turns out to be equivalent to the theory of shallow water equations, and
a particular example is the one-dimensional system of Benney long wave equations, which is
integrable \cite{Be1973, Gi1981}.

Here we shall consider the singular
solutions of the geodesic dynamics of the Vlasov $p-$moments. Remarkably,
these equations turn out to be related to other integrable systems governing shallow water
wave theory. For example, when the Vlasov $p-$moment equations for geodesic
motion on the symplectomorphisms are closed at the level of the first
$p-$moment, their singular solutions are found to recover the peaked soliton of the
integrable Camassa-Holm (CH) equation for shallow water waves \cite{CaHo1993}. 
These singular Vlasov moment solutions also correspond to individual particle motion. 

Thus, geodesic symplectic dynamics of the Vlasov $p-$moments is found to
possess singular solutions whose closure at the fluid level for the CH equation
recovers the
peakon solutions of shallow water theory. Being solitons, the CH peakons
superpose and undergo elastic collisions in fully nonlinear interactions.
The singular solutions for Vlasov $p-$moments presented here also
superpose and interact nonlinearly as coherent structures. \\
\rem{
However, they
differ qualitatively from the singular solutions of both the Camassa-Holm
equation and the Vlasov equation. This difference arises because the
momentum of the $j-$th singular solution for the $p-$moments is a
function of its corresponding canonically conjugate single-particle
momentum, as well as being a function of time.} 

\noindent
The plan of the paper follows: 
%\todo{Revise}
\begin{description}
\item
Section \ref{pmom-eqns} reviews the Vlasov $p-$moment equations and
recounts their Lie-Poisson Hamiltonian structure using the Kupershmidt-Manin
Lie-Poisson bracket. Variational formulations of the
$p-$moment dynamics are also provided.
%\item
%Section \ref{VarPrincHP} derives variational formulations of the
%$p-$moment dynamics in both their Lagrangian and Hamiltonian forms. 
\item
Section \ref{colifts} shows how the Lagrangian framework for fluid dynamics
is recovered from the Vlasov $p-$moments and establishes connections with some equations for shallow water waves. In this case, the $p-$moment equations are
shown to possess singular solutions. 
\item
Section \ref{Apps} establishes the connections between the integrable Benney equations and  the physics of charged-particle accelerator beams. To our knowledge, these connections are noted here for the first time. We also point out how the experimental realization
of solitary waves in coasting particle beams has its roots in the integrability of
the Benney system. 
\rem{  % --------- BEGIN REM
The integrable EPDiff and Camassa-Holm equations are also
presented as other important applications of moment dynamics. Remarkably, these integrable equations also possess singular solutions. An important feature shared among these systems seems to be that both for applications in physics and for integrable systems the moment Hamiltonians are quadratic.
}% --------- END REM
\item
Section \ref{geo-prob} formulates the problem of geodesic motion on the
symplectomorphisms in terms of the Vlasov $p-$moments and identifies the
singular solutions of this problem. This geodesic motion
is related to a geodesic form of the Vlasov equation. Thus the singular solutions are found to originate in the single particle dynamics on phase space. In a special case, the truncation of geodesic symplectic motion to geodesic diffeomorphic motion for
the first $p-$moment recovers the singular solutions of the Camassa-Holm
equation, and thereby correspond to single particle dynamics.
\rem{
\item
Section \ref{geo-prob} discusses how the singular $p-$moment solutions
for geodesic symplectic motion are related to the cold plasma solutions.
By symmetry under exchange of canonical momentum $p$ and position $q$, the
Vlasov $q-$moments are also found to admit singular (weak) solutions.
}
\end{description}

The geodesic form of the Vlasov equation was introduced in \cite{GiHoTr05},
where it was also shown how to extend the treatment to higher dimensions.

\section{Review of Vlasov moment dynamics}\label{pmom-eqns}

The Vlasov equation may be expressed as 
\begin{eqnarray}
\frac{\partial f}{\partial t}
=
\Big[f\,,\,\frac{\delta h}{\delta f}\Big]
=
\frac{\partial f}{\partial p}\frac{\partial}{\partial q}
\frac{\delta h}{\delta f}
-
\frac{\partial f}{\partial q}\frac{\partial}{\partial p}
\frac{\delta h}{\delta f}
=: -\,{\rm ad}^*_{\delta h/\delta f}\,f
\label{vlasov-eqn}
\end{eqnarray}
Here the canonical Poisson bracket $[\,\cdot\,,\,\cdot\,]$
is defined for smooth functions on phase space with
coordinates $(q,p)$ and $f(q,p,t)$ is the evolving Vlasov
single-particle distribution function.  The variational
derivative $\delta h/\delta f$ is the single
particle Hamiltonian and the ${\rm ad}^*_{\delta h/\delta f}\,f$ is explained
as follows.

A functional $g[f]$ of the Vlasov distribution $f$ evolves
according to 
\begin{eqnarray*}
\frac{dg}{dt}
&=&
\int\hspace{-3mm}\int
\frac{\delta g}{\delta f}\,
 \frac{\partial f}{\partial t} 
\,dqdp 
=
\int\hspace{-3mm}\int
\frac{\delta g}{\delta f}\,
\Big[f\,,\,\frac{\delta h}{\delta f}\Big]
\,dqdp 
\\
&=&
-\int\hspace{-3mm}\int f 
\Big[\frac{\delta g}{\delta f}\,,\,\frac{\delta h}{\delta f}\Big]
\,dqdp 
=:-\,
\Big\langle\!\!\Big\langle f\,,\,
\Big[\frac{\delta g}{\delta f}\,,\,\frac{\delta h}{\delta f}\Big]
\Big\rangle\!\!\Big\rangle
=:
\{\,g\,,\,h\,\}
\end{eqnarray*}
In this calculation boundary terms were neglected upon integrating by
parts in the third step and the notation
$\langle\!\langle\,\cdot\,,\,\cdot\,\rangle\!\rangle$ is introduced for
the $L^2$ pairing in phase space. The quantity $\{\,g\,,\,h\,\}$ defined
in terms of this pairing is the Lie-Poisson Vlasov (LPV) bracket
\cite{WeMo}. This Hamiltonian evolution equation may also be expressed as
\begin{eqnarray*}
\frac{dg}{dt}
=
\{\,g\,,\,h\,\}
=-
\Big\langle\!\!\Big\langle f\,,\,
{\rm ad}\,_{\delta h/\delta f}
\frac{\delta g}{\delta f}
\Big\rangle\!\!\Big\rangle
=
-\,
\Big\langle\!\!\Big\langle 
{\rm ad}^*\,_{\delta h/\delta f}\,f\,,\,
\frac{\delta g}{\delta f}
\Big\rangle\!\!\Big\rangle
\end{eqnarray*}
which defines the Lie-algebraic operations ad and ad$^*$ in this
case in terms of the $L^2$ pairing on phase space
$\langle\!\langle\,\cdot\,,\,\cdot\,\rangle\!\rangle$:
$\mathfrak{s}^*\times\mathfrak{s}\mapsto\mathbb{R}$. The
notation ${\rm ad}^*_{\delta h/\delta f}\,f$ in
(\ref{vlasov-eqn}) expresses {\bfi coadjoint action} of $\delta
h/\delta f\in\mathfrak{s}$ on $f\in\mathfrak{s}^*$, where
$\mathfrak{s}$ is the Lie algebra of single particle Hamiltonian
vector fields and $\mathfrak{s}^*$ is its dual under $L^2$
pairing in phase space. This is the sense in which the Vlasov equation
represents coadjoint motion on the symplectomorphisms. This Lie-Poisson structure
has also been extended to include Yang-Mills theories in \cite{GiHoKu1982}
and \cite{GiHoKu1983}.

In higher dimensions, particularly $n = 3$, we may take the direct sum of the Vlasov Lie-Poisson bracket, together with with the Poisson bracket for an electromagnetic field (in the Coulomb gauge) where the electric field $\bf E$ and magnetic vector potential $\bf A$ are canonically conjugate.
For discussions of the Vlasov-Maxwell equations from a geometric viewpoint in the same spirit as the present approach, see \cite{CeHoHoMa1998}, \cite{MaWe},
\cite{MaWeRaScSp} and \cite{WeMo}.

\subsection{Dynamics of Vlasov $q,p-$Moments}
The phase space $q,p-$moments of the Vlasov distribution function are
defined by 
\begin{eqnarray*}
g_{\,\widehat{m}m}
&=&
\int\hspace{-3mm}\int f (q,p)\,
q^{\widehat{m}}p^m
\,dq\,dp 
\,.
\end{eqnarray*}
The $q,p-$moments $g_{\,\widehat{m}m}$ are often used in
treating the collisionless dynamics of plasmas and particle beams
\cite{Dragt}. This is usually done by considering low-order
truncations of the potentially infinite sum over phase space
moments,  
\begin{eqnarray*}
g
&=&
\sum_{\widehat{m},m=0}^\infty
a_{\,\widehat{m}m}g_{\,\widehat{m},m}
\,,\qquad
h
=
\sum_{\widehat{n},n=0}^\infty
b_{\,\widehat{n}n}g_{\,\widehat{n},n}
\,,
\end{eqnarray*}
with constants $a_{\,\widehat{m}m}$ and $b_{\,\widehat{n}n}$,
with $\widehat{m},m,\widehat{n},n=0,1,\dots$. If $h$ is the
Hamiltonian, the sum over $q,p-$moments $g$ evolves under the Vlasov
dynamics according to the Poisson bracket relation
\begin{eqnarray*}
\frac{dg}{dt}
=
\{\,g\,,\,h\,\}
&=&
\sum_{\widehat{m},m,\widehat{n},n=0}^\infty
a_{\,\widehat{m}m}b_{\,\widehat{n}n}
(\widehat{m}m-\widehat{n}n)
g_{\,\widehat{m}+\widehat{n}-1,m+n-1}
\,.
\end{eqnarray*}
The symplectic invariants associated with Hamiltonian
flows of the $q,p-$moments were discovered and classified
in \cite{HoLySc1990}. Finite dimensional approximations of the whole $q,p-$moment
hierarchy were discussed in \cite{ScWe}. For 
discussions of the Lie-algebraic approach to the control and  steering of
charged particle beams, see 
\cite{Dragt}.

\subsection{Dynamics of Vlasov $p-$Moments} 

In contrast to the $q,p-$moments, the momentum moments, or ``$p-$moments,'' of the Vlasov function are
defined as
\begin{eqnarray*}
A_m(q,t)=\int p^m\,f(q,p,t)\,dp
\,,\qquad
m=0,1,\dots.
\end{eqnarray*}
 That is, the $p-$moments are $q-$dependent integrals over $p$
 of the product of powers $p^m$, $m=0,1,\dots$, times the
 Vlasov solution $f(q,p,t)$.
We shall consider functionals of these $p-$moments defined by,
\begin{eqnarray*}
g
&=&
\sum_{m=0}^\infty
\int\hspace{-3mm}\int
 \alpha_m(q)\,p^m\,f\,dqdp
=
\sum_{m=0}^\infty
\int
 \alpha_m(q)\,A_m(q)\,dq
=:
\sum_{m=0}^\infty\Big\langle A_m\,,\,\alpha_m\Big\rangle
\,,
\\
h
&=&
\sum_{n=0}^\infty
\int\hspace{-3mm}\int
 \beta_n(q)\,p^n\,f\,dqdp
=
\sum_{n=0}^\infty
\int
 \beta_n(q)\,A_n(q)\,dq
=:
\sum_{n=0}^\infty\Big\langle A_n\,,\,\beta_n\Big\rangle
\,,
\end{eqnarray*}
where $\langle\,\cdot\,,\,\cdot\,\rangle$ is the
$L^2$ pairing on position space.

The functions $\alpha_m$
and $\beta_n$ with $m,n=0,1,\dots$ are assumed to be suitably
smooth and integrable against the Vlasov $p-$moments. To
assure these properties, one may relate the $p-$moments to the
previous sums of Vlasov $q,p-$moments by choosing
\begin{eqnarray*}
\alpha_m(q)
&=&
\sum_{\widehat{m}=0}^\infty
a_{\,\widehat{m}m}q^{\,\widehat{m}}
\qquad\text{ and }\qquad
\beta_n(q)
=
\sum_{\widehat{n}=0}^\infty
b_{\,\widehat{n}n}q^{\,\widehat{n}}
\,.
\end{eqnarray*}
For these choices of $\alpha_m(q)$ and $\beta_n(q)$, the sums of 
$p-$moments will recover the full set of Vlasov $(q,p)-$moments.
Thus, as long as the $q,p-$moments of the distribution $f(q,p)$
continue to exist under the Vlasov evolution, one may assume that the
dual variables $\alpha_m(q)$ and $\beta_n(q)$ are smooth functions whose
Taylor series expands the $p-$moments in the $q,p-$moments. These functions
are dual to the $p-$moments $A_m(q)$ with $m=0,1,\dots$ under the $L^2$
pairing $\langle\cdot\,,\,\cdot\rangle$ in the spatial variable $q$.
In what follows we will assume {\it homogeneous} boundary conditions. This means, for example, that we will ignore boundary terms arising from integrations by parts.

%\subsection{Poisson bracket for Vlasov $p-$moments}

The Poisson bracket among the
$p-$moments is obtained from the LPV bracket 
\rem{ % -------BEGIN REM
via the following
explicit calculation, 
\begin{eqnarray*}
\{\,g\,,\,h\,\}
&=&
-\sum_{m,n=0}^\infty
\int\hspace{-3mm}\int f 
\Big[\alpha_m(q)\,p^m\,,\,\beta_n(q)\,p^n\Big]
\,dqdp 
\\
&=&
-\sum_{m,n=0}^\infty
\int\hspace{-3mm}\int 
\Big[
m\alpha_{m}\beta_{n}\,^{\prime}\,-n\beta_{n}\alpha_{m}\,^{\prime}\Big]
f\,p^{m+n-1}
\,dqdp 
\\
&=&
-\sum_{m,n=0}^\infty
\int
A_{m+n-1}(q)
\Big[
m\alpha_{m}\beta_{n}\,^{\prime}\,-n\beta_{n}\alpha_{m}\,^{\prime}\Big]
\,dq
\\
&=:&
\sum_{m,n=0}^\infty\Big\langle 
A_{m+n-1}
\,,\,
{\rm ad}_{\beta_n}\alpha_m
\Big\rangle 
\\
&=&
-\sum_{m,n=0}^\infty
\int
\Big[
n\beta_n A_{m+n-1}'
+(m+n)A_{m+n-1}\beta_n\,'\Big]
\alpha_m
\,dq
\\
&=:&
-\sum_{m,n=0}^\infty\Big\langle 
{\rm ad}^*_{\beta_n}A_{m+n-1}
\,,\,
\alpha_m
\Big\rangle 
\end{eqnarray*}
where we have integrated by parts and introduced the 
notation ad and ad$^*$ for adjoint and coadjoint action,
respectively. Upon recalling the dual relations 
\begin{eqnarray*}
\alpha_m=\frac{\delta g}{\delta A_m}
\quad\hbox{and}\quad
\beta_n=\frac{\delta h}{\delta A_n}
\end{eqnarray*}
the LPV bracket in terms of the $p-$moments 
}
may be expressed as
\begin{eqnarray}
\{\,g\,,\,h\,\}(\{A\})
&=&
-\sum_{m,n=0}^\infty
\int
\frac{\delta g}{\delta A_m}
\Big[
n\frac{\delta h}{\delta A_n} 
\frac{\partial}{\partial q} A_{m+n-1}
+
(m+n)A_{m+n-1}\frac{\partial}{\partial q}
\frac{\delta h}{\delta A_n}\Big]
\,dq
\nonumber \\
&=:&
-\sum_{m,n=0}^\infty
\Big\langle 
A_{m+n-1}
\,,\,
\Big[\!\!\Big[\frac{\delta g}{\delta A_m}\,,\,
\frac{\delta h}{\delta A_n}\Big]\!\!\Big]
\Big\rangle 
\label{KMLP-brkt}
\end{eqnarray}
This is the Kupershmidt-Manin Lie-Poisson (KMLP) bracket
\cite{KuMa}, which is defined for
functions on the dual of the Lie algebra with bracket
\begin{eqnarray*}
[\![\,\alpha_m\,,\,\beta_n\,]\!]
=
m\alpha_m\partial_q\beta_n-n\beta_n\partial_q\alpha_m
\,.
\end{eqnarray*}
This Lie algebra bracket inherits the Jacobi identity
from its definition in terms of the canonical Hamiltonian
vector fields. Thus, we have shown the
\begin{theorem}{\large\bf (Gibbons \cite{Gi1981})}$\,$
The operation of taking
$p-$moments of Vlasov solutions is a Poisson map. It takes the LPV bracket
describing the evolution of $f(q,p)$ into the KMLP bracket, describing the
evolution of the $p-$moments $A_n(x)$. 
\end{theorem}
A result related to this, for
the Benney hierarchy \cite{Be1973}, was also noted by Lebedev and Manin
\cite{LeMa}.

The evolution of a particular $p-$moment $A_m(q,t)$ is obtained
from the KMLP bracket by
\begin{align*}
\frac{\partial  A_m}{\partial t}
\,=\,
\{\,A_m\,,\,h\,\}
&\,=\,
-\sum_{n=0}^\infty
\Big(n\frac{\partial}{\partial q} A_{m+n-1}
+
mA_{m+n-1}\frac{\partial}{\partial q}
\Big)\,
\frac{\delta h}{\delta A_n}
\,=\,
\sum_{n=0}^{\infty}\{\,A_{m}\,,\,A_{n}
\,\}\,\frac{\delta h}{\delta A_{n}}
\end{align*}

These moment equations can also be derived from variational principles, as
shown in \cite{GiHoTr05}, within the Hamilton-Poincar\'e framework \cite{CeMaPeRa}.

\section{Moments and cotangent lifts of diffeomorphisms}\label{colifts}
As explained in the introduction, a first order closure of the moment hierarchy leads to the
equations of ideal fluid dynamics. Such equations represent coadjoint motion with respect to the Lie group of smooth
invertible maps (diffeomorphisms). This coadjoint evolution may be interpreted in terms of Lagrangian variables, which are invariant under the action of diffeomorphisms. In this section we investigate how the entire moment hierarchy may be expressed in terms of the fluid quantities evolving
under the diffeomorphisms and express the conservation laws in this case.

\subsection{Lagrangian variables and cotangent lifts}
In order to look for Lagrangian variables, we consider the geometric interpretation
of the moments, regarded as \emph{fiber integrals} on the cotangent bundle
$T^*Q$ of some configuration manifold $Q$. A $p-$moment is defined as a fiber
integral; that is, an integral on the single
fiber $T^*_q Q$ with base point $q\in Q$ kept fixed
\[
A_n(q)=\int_{T^*_q Q}\, p^n \,f(q,p)\, dp
\]
A similar approach is followed for gyrokinetics in \cite{QiTa2004}.
Now, 
\rem{
consider a characteristic curve given by the action of canonical transformations:
\[
A_n^{(t)}(q_t)=\int_{T^*_{q_t}\! Q}\, \left[p_t(q_0,p_0)\right]^n \,\,f_{t}\!\left(q_t(q_0,p_0),\,p_t(q_0,p_0)\right)\,\, dp_t(q_0,p_0)
\]
}
the problem is that in general the integrand does not stay on one same fiber
under the action of canonical transformations,
i.e. symplectomorphisms are not \emph{fiber-preserving}
in the general case. However, one may avoid this problem by restricting to a subgroup of these canonical
transformations whose action is fiber preserving. The transformations in this subgroup (indicated with $T^*\text{Diff}(Q)$) are called \emph{point transformations} or \emph{cotangent lifts} of diffeomorphisms  and they arise from diffeomorphisms on points in configuration space \cite{MaRa99}, such that 
\[
q_t=q_t(q_{_0})
\]
The fiber preserving nature of cotangent lifts is expressed by the preservation
of the canonical one-form:
\[
p_t\, dq_t=p_t(q_{_0},p_{_0})\, dq_t(q_{_0})=p_{_0} dq_{_0}
\]
This fact also reflects in the particular form assumed by the generating
functions of cotangent lifts, which are linear in the momentum coordinate,
i.e.
\[
H(q,p)\,=\,\beta(q)\,\frac{\partial}{\partial q}\contract\, p\,dq\,=\,p\,\beta(q)\,.
\]
Restricting to cotangent lifts represents a limitation in comparison with considering the whole symplectic group. However, this is a natural
way of recovering the fluid equations, starting from the full moment dynamics.

\subsection{Characteristic equations for the moments}
Once one restricts to cotangent lifts, Lagrangian moment variables may be defined
and conservation laws may be found, as in the context of fluid
dynamics. The key idea is to use the preservation of the canonical one-form
for constructing invariant quantities. Indeed one may take $n$ times
the tensor product of the canonical one-form with itself and write:
\[
p_t^{\,n}\left(dq_t\right)^n=p_0^{\,n}\left(dq_0\right)^n
\]
\rem{
where
\[
\left(dq\right)^n \,=\,
\underset{n \text{ times}}{\underbrace{dq\otimes\dots\otimes dq}}
\]
}
One then considers the preservation of the Vlasov density
\[
f_t(q_t,p_t) \,dq_t\wedge dp_t = f_0(q_0,p_0)\, dq_0\wedge dp_0
\]
and writes
\[
p_t^{\,n} \,f_t(q_t,p_t) \,(dq_t)^n\otimes dq_t\wedge dp_t=
p_0^{\,n}\,f_0(q_0,p_0)\, (dq_0)^n\otimes dq_0\wedge dp_0
\]
Integration over the canonical particle momenta yields the following characteristic equations
\begin{equation}\label{conslaw}
\frac{d}{dt}\!\left[A^{(t)}_n(q_t)\,(dq_t)^n\otimes dq_t\right] = 0\qquad\text{along}\quad
\dot{q_t}=\frac{\partial H}{\partial p}=\beta(q)
\end{equation}
\rem{
where the single-particle Hamiltonian belongs now to the Lie algebra of cotangent
lifts and is thus written as
\[
h(q,p)=p\,u(q)
\]
with $u(q)$ being a vector field on $Q$, so that $\dot{q_t}=u(q)$.
}
which recover the well known conservations for fluid density and momentum ($n=0,1$) and can be
equivalently written in terms of the Lie-Poisson equations arising from the KMLP bracket. Indeed, if the vector field $\beta$ is identified with the Lie algebra variable $\beta=\beta_1=\delta h/\delta A_1$ and $h(A_1)$ is the moment Hamiltonian, then
the KMLP form of the moment equations is
\[
\dot{A}_n+\textrm{\large ad}^*_{\beta_1} A_n=0.
\]
In this case, the KM $\textrm{ad}^*_{\beta_1}$ operation coincides with the Lie derivation $\pounds_{\!\beta}$. Hence, one may also write it equivalently
as
\[
\dot{A}_n+{\pounds}_{\!\beta} \,A_n=0.
\]
This equation is reminiscent of the so called {\it b-equation} introduced in \cite{HoSt03}, for which the vector field $\beta$ is nonlocal and may
be taken as $\beta(q)=G*A_n$ (for any $n$), where $G$ is the Green's function of the Helmholtz operator and the star denotes convolution.
%\todo{We probably need more remarks on integrability for particular choices
%of $b$ (or $n$)}
When the vector field $\beta$ is sufficiently smooth, this equation
is known to possess singular solutions of the form
\[
A_n(q,t)=\sum_{i=1}^K P_{n,\,i}(t)\,\delta(q-Q_{i}(t))
\]
where the $i-$th position $Q_i$ and weight $P_{n,\,i}$ of the singular solution for the $n-$th moment satisfy the following equations
\begin{align*}
\dot Q_i&=\left.\beta(q)\right|_{q=Q_i}\\
\dot P_{n,\,i}&=-\,n\, P_{n,\,i}\left.\frac{\partial \beta(q)}{\partial q}\right|_{q=Q_i}
\end{align*}
Interestingly,
for $n=1$ (with $\beta(q)=G*A_1$), these equations recover the pulson solutions of the Camassa-Holm equation, which play an important role in the following discussion.
Moreover the particular case $n=1$ represents the single particle solution
of the Vlasov equation. However, when $n\neq1$ the interpretation of these
solutions as single-particle motion requires the particular
choice $P_{n,\,i}=(P_i)^{\,n}$. For this choice, the $n-$th weight is identified with the $n-$th power of the particle momentum. 

The equations presented in this section provide a geometric interpretation of the moments in terms of the covariant tensor power densities in (\ref{conslaw}). When one restricts to cotangent lifts, one can specify an action of the group of diffeomorphisms on the moments. However, this action is not yet understood at the group level in the general case. The general geometric nature of the Lie algebra would be expressed in terms of contravariant tensors dual to the covariant tensor power densities. For the cotangent lifts, these contravariant tensors reduce to contravariant vector fields. In this case, one is able to characterize their action on the moments as Lie derivatives.

\paragraph{KMLP bracket and cotangent lifts.}
We have seen that restricting to cotangent lifts
leads to a Lagrangian fluid-like formulation of the dynamics of the resulting
$p$-moments. In this case, the moment
equations are given by the KMLP bracket when the Hamiltonian depends only
on the first moment ($\beta_1=\delta h/\delta A_1$)
\[
\{g,h\}=-\sum_n
\bigg\langle 
A_{n}
\,,\,
\bigg[\!\!\bigg[\frac{\delta g}{\delta A_n}\,,\,
\frac{\delta h}{\delta A_1}\bigg]\!\!\bigg]
\bigg\rangle 
\]
If one now restricts the bracket to functionals of only the first moments,
one may check that the KMLP bracket yields the well known Lie-Poisson bracket
on the group of diffeomorphisms
\[
\{g,h\}[A_1]=\left\langle A_1,\,\left[
\frac{\delta h}{\delta A_1}\,
\frac{\partial}{\partial q}\frac{\delta g}{\delta A_1}
- 
\frac{\delta g}{\delta A_1}\,
\frac{\partial}{\partial q}\frac{\delta h}{\delta A_1}
\right]\,\right\rangle
\]
This is a very natural step since diffeomorphisms
and their cotangent lifts are isomorphic. In fact, this is the bracket used for ideal incompressible fluids as well as for the construction of the EPDiff equation, which
will be discussed later as an application of moment dynamics.

\section{Applications of the KMLP bracket and quadratic terms}\label{Apps} 

\subsection{The Benney equations}
The KMLP bracket (\ref{KMLP-brkt}) was first derived in the context of Benney long waves,
whose Hamiltonian is
\begin{equation*}
H=\frac12\int (A_2(q)+gA_0^2(q))\,dq.
\end{equation*}
The Hamiltonian form $\partial_tA_n=\{A_n,\,H\}$ with the KMLP bracket leads to the moment equations
\begin{equation*}
\frac{\partial A_n}{\partial t}
+
\frac{\partial A_{n+1}}{\partial q}
+g
n A_{n-1}
\frac{\partial A_0}{\partial q}
=0
\end{equation*}
derived by Benney \cite{Be1973} as a description of long waves on a
shallow perfect fluid, with a free surface at $y=h(q,t)$. In his
interpretation, the $A_n$ were vertical moments of the horizontal
component of the velocity $p(q,y,t)$:
\begin{equation*}
A_n=\int_{0}^{h} p^n(q,y,t) \,\text{d}y.
\end{equation*}
The corresponding system of evolution equations for $p(q,y,t)$ and $h(q,t)$
is related by hodograph transformation, $y=\int_{-\infty}^p f(q,p',t)\,
\text{d}p'$, to the Vlasov equation
\begin{equation}\label{Vlasov-Benney}
\frac{\partial f}{\partial t}
+
p
\frac{\partial f}{\partial q}
-g
\frac{\partial A_0}{\partial q}
\frac{\partial f}{\partial p}
=0.
\end{equation}
The most important fact about the Benney hierarchy is that it is completely
integrable. This fact emerges from the following observation. Upon defining a
function $\lambda(q,p,t)$ by the principal value integral, 
\begin{equation*}
\lambda(q,p,t)=
p+P\int_{-\infty}^\infty \frac{f(q,p\,',t)}{p-p\,'} \,\text{d}p\,',
\end{equation*}
it is straightforward to verify \cite{LeMa} that
\begin{equation*}
\frac{\partial \lambda}{\partial t}
+
p
\frac{\partial \lambda}{\partial q}
-g
\frac{\partial A_0}{\partial q}
\frac{\partial \lambda}{\partial p}
=0;
\end{equation*}
so that $f$ and $\lambda$ are advected along the same characteristics.

\rem{
In higher dimensions, particularly $n=3$, we may take the direct sum of the KMLP bracket, together with with the Poisson bracket for an electromagnetic field (in the Coulomb gauge) where the electric field $\mathbf{E}$ and magnetic vector potential $\mathbf{A}$ are canonically conjugate; then the Hamiltonian

\begin{equation*} 
H_{MV} =\int \int \left[\frac{1}{2m} |\mathbf{p} - e\mathbf{A}|^2 \right] f(\mathbf{p},\mathbf{q}) \mathrm{d}^n \mathbf{p} \mathrm{d}^n \mathbf{q} + \int \left[\frac{1}{2} |\mathbf{E}|^2 +  \frac{1}{4}\sum_{i=1}^n
\sum_{j=1}^n (A_{i,j}-A_{j,i})^2 \right]   \mathrm{d}^n \mathbf{q} 
\end{equation*} 
yields the Maxwell-Vlasov (MV) equations for systems of interacting
charged particles. For a discussion of the MV equations from a geometric
viewpoint in the same spirit as the present approach, see \cite{CeHoHoMa1998}.
}

\paragraph{Applications to coasting accelerator beams.}
Interestingly, the Vlasov equation (\ref{Vlasov-Benney}) resulting from the
hodograph transformation of the Benney equation is exactly the same
as the equation that regulates coasting proton beams in particle accelerators
(see for example \cite{Venturini} where a bunching term is also included).\smallskip

\noindent
Now, this is a very remarkable fact: the integrability of the \emph{Vlasov-Benney}
equation implies soliton solutions, indications for which seem to have
been found experimentally at CERN
%the CERN Proton Synchrotron Booster
\cite{KoHaLi2001}, 
%at the Brookhaven Alternating Gradient Synchrotron 
BNL \cite{BlBrGlRaRy03}, LANL \cite{CoDaHoMa04} and 
%at the Fermilab Tevatron 
FermiLab \cite{MoBaJaLeNgShTa}. (In the last case solitons are shown to appear even when a bunching force is present.) These solutions have attracted the attention of the accelerator community and considerable analytical work has been carried out over the last decade (see
for example \cite{ScFe2000}). Nevertheless to our knowledge, the existence of solitons in coasting proton beams has never been related to the integrability of the governing Vlasov equation via its connection to the Benney hierarchy,, although this would explain very naturally why robust coherent structures are seen in these experiments as fully nonlinear excitations.  We plan to pursue this direction in future research. 

\rem{
Rather, connections with the well known integrable KdV equation have been proposed \cite{ScFe2000}, but we believe
this is not a natural step since integrability appears directly in the Vlasov-Benney system that governs the collective motion of the beam.
}

\subsection{The Vlasov-Poisson system, the wakefield model and singular solutions}

Besides integrability of the Vlasov-Benney equation, there are other important
applications of the Vlasov equation that have in common the presence of
a quadratic term in $A_0$ within the Hamiltonian:
\[
H=\frac12\int\! A_2(q)\,dq+\frac12\iint\! A_0(q)\,G(q,q')\,A_0(q')\,dq\,dq'.
\]
For example, when $G=\left(\partial_q^{\,2}\right)^{-1}$, this Hamiltonian leads to the Vlasov-Poisson
system, which is of fundamental importance in many areas of plasma physics.
More generally, this Hamiltonian is widely used for beam dynamics in particle
accelerators: in this case $G$ is related to the electromagnetic interaction of a beam with the vacuum chamber. The {\bfi wake field} is originated by the image charges induced on the walls by the passage of a moving particle: while the beam passes, the charges in the walls are attracted towards the inner surfaces and generate a field that acts back on the beam. This affects the dynamics of the beam, thereby causing several problems such as beam energy spread and instabilities. In the literature, the {\bfi wake function} $W$ is introduced so that \cite{Venturini}
\[
G(q,q')=\int_{-\infty}^q \!W(x, q')\,dx
\]
Wake functions usually depend only on the properties of the accelerator chamber.

An interesting wakefield model has been presented in \cite{ScFe2000}
where $G$ is chosen to be the Green's function of the Helmholtz operator
$\left(1-\alpha^2\partial_q^2\right)$: this generates a \emph{Vlasov-Helmholtz} (VH) equation
\cite{CaMaPu2002} that is particularly interesting for future work. Connections of this equation with the well known integrable KdV equation have been proposed. However we believe
that
this is not a natural step since integrability appears already with no further
approximations in the Vlasov-Benney system that governs the collective motion of the beam. In particular we would
like to understand the VH equation as a special deformation of the integrable
Vlasov-Benney case that allows the existence of {\bfi singular solutions}.
%\comment{A few more lines added on singular solutions.}
 Indeed,
the presence of the Green's function $G$ above is a key
ingredient for the existence of the single-particle solution, which is \emph{not}
allowed in the VB case. 
\rem{
In particular, the single-particle solutions for the VH equation might be of great interest, since they are singular solutions that arise from a deformation of an integrable system
and whose properties can be very different from the well known particle behaviour
of the Vlasov-Poisson case. Varying the length-scale $\alpha$ in the Helmholtz
Green's function would be a
crucial path when approaching the integrable limit $\alpha\rightarrow0$.
}
In particular, the single-particle solutions for the Vlasov-Helmholtz equation may be of great interest, since these singular solutions arise from a deformation of an integrable system. The Vlasov-Helmholtz solutions may differ considerably from the well known particle behavior of the Vlasov-Poisson case. However, the limit as the deformation parameter (the length-scale $\alpha$) in the Helmholtz Green's function passes to zero ($\alpha\rightarrow0$) in the wake-field equations, one recovers the integrable Vlasov-Benney case.

\subsection{The EPDiff equation and singular solutions}
Another interesting  moment equation is given by the integrable EPDiff equation \cite{CaHo1993}. In this case, the Hamiltonian is purely quadratic in the first moments:
\[
H=\frac12\iint \!A_1(q)\, G(q,q')\, A_1(q')\, dq\,dq'
\]
and the EPDiff equation \cite{HoMa2004}
\[
\frac{\partial A_{1}}{\partial t} 
+\frac{\partial A_{1}}{\partial q}\int \!G(q,q') A_{1}(q',t)dq'
+2A_{1}\,\frac{\partial }{\partial q}\int \!G(q,q') A_{1}(q',t)dq'=0
\]
comes from the closure of the KMLP bracket given by cotangent lifts. (Without
this restriction we would obtain again the equations (\ref{conslaw}) with
$\beta=G* A_1$.) Thus this EPDiff equation is a \emph{geodesic} equation on the group
of diffeomorphisms. The Camassa-Holm equation is a particular case in which $G$ is the Green's function of the Helmholtz operator $1-\alpha^2\partial_q^2$. Both the CH and the EPDiff equations are completely
 integrable and have a
large number of applications in fluid dynamics (shallow water theory, averaged
fluid models, etc.) and imaging techniques \cite{HoRaTrYo2004} (medical imaging, contour dynamics,
etc.). 

Besides its complete integrability the EPDiff equation has the important feature of allowing singular delta-function solutions. The connection between the CH and EPDiff equations and the moment dynamics lies in the fact that  singular solutions appear in both contexts.
The existence of this kind of solution for EPDiff leads to investigate its origin in the context of Vlasov moments. More particularly we wonder whether there is a natural extension of the EPDiff equation to all the moments. This would again be a geodesic
(hierarchy of) equation, which would perhaps explain how the singular solutions for EPDiff arise in this larger context.

\section{Geodesic motion and singular solutions}\label{geo-prob}
\subsection{Quadratic Hamiltonians}
The previous examples show how quadratic terms in the Hamiltonian produce
interesting behaviour in
various contexts. This suggests that a deeper analysis of the role of quadratic
terms may
be worthwhile particularly in connections between Vlasov $p-$moment dynamics
and the EPDiff equation, with its singular solutions.
Purely quadratic Hamiltonians are considered  in \cite{GiHoTr05},
leading to the problem of geodesic motion on the space of
$p-$moments. 

we are interested in the problem of geodesic motion on the space of
$p-$moments. In this problem the Hamiltonian is the norm on the
$p-$moment given by the following metric and inner product,
\begin{eqnarray}
h=\frac{1}{2}\|A\|^2
&=&
\frac{1}{2}\sum_{n,s=0}^\infty
\int\hspace{-3mm}\int
A_n(q)G_{ns}(q,q\,')A_s(q\,')\,dq\,dq\,'
\label{Ham-metric}
\end{eqnarray}
The metric $G_{ns}(q,q\,')$ in (\ref{Ham-metric}) is chosen to be positive definite,
so it defines a norm for $\{A\}\in\mathfrak{g}^*$. The
corresponding geodesic equation with respect to this norm is
found as in the previous section to be,
\begin{eqnarray}
\frac{\partial  A_m}{\partial t}
=
\{\,A_m\,,\,h\,\}
=
-\sum_{n=0}^\infty
\Big(n\beta_n
\frac{\partial}{\partial q} A_{m+n-1}
+
(m+n)A_{m+n-1}\frac{\partial}{\partial q}
\beta_n
\Big)
\label{EPMS-eqn}
\end{eqnarray}
with dual variables $\beta_n\in\mathfrak{g}$ defined by
\begin{eqnarray}
\beta_n
=
\frac{\delta h}{\delta A_n} 
=
\sum_{s=0}^\infty
\int
G_{ns}(q,q\,')A_s(q\,')\,dq\,'
=
\sum_{s=0}^\infty
G_{ns}*A_s
\,.
\label{EPMS-vel}
\end{eqnarray}
Thus, evolution under (\ref{EPMS-eqn}) may be rewritten as
coadjoint motion on $\mathfrak{g}^*$
\begin{eqnarray}
\frac{\partial  A_m}{\partial t}
=
\{\,A_m\,,\,h\,\}
=:
-\sum_{n=0}^\infty
{\rm ad}^*_{\beta_n}A_{m+n-1}
\label{A-dot}
\end{eqnarray}
This system comprises an infinite system of nonlinear, nonlocal, coupled
evolutionary equations for the $p-$moments. In this system, evolution of
the $m^{th}$ moment is governed by the potentially infinite sum of
contributions of the velocities $\beta_n$ associated with $n^{th}$
moment sweeping the $(m+n-1)^{th}$ moment by a type of coadjoint action.
Moreover, by equation (\ref{EPMS-vel}), each of the $\beta_n$
potentially depends nonlocally on all of the moments. 

\rem{
Equations (\ref{Ham-metric}) and (\ref{EPMS-vel}) may be
written in three dimensions in multi-index notation, as follows:
the Hamiltonian is given by
\[
h
=
\frac{1}{2}
\left\vert 
\left\vert 
A
\right\vert 
\right\vert 
^{2}
=
\frac{1}{2}
\sum\limits_{\mu,\nu}
\iint
A_{\mu}
\left(  
\mathbf{q},
t
\right)  
G_{\mu\nu}
\left(  
\mathbf{q,q}\,^{\prime}
\right)  
A_{\nu}
\left(  
\mathbf{q}\,^{\prime},
t
\right)  
d
\mathbf{q}
d
\mathbf{q}\,^{\prime}
\]
so the dual variable is written as
\begin{align*}
\beta_{\rho}
=
\frac{\delta h}{\delta A_{\rho}}  
& =
\sum\limits_{\nu}
\iint
G_{\rho\nu}
\left(  
\mathbf{q,q}\,^{\prime}
\right)  
A_{\nu}
\left(  
\mathbf{q}\,^{\prime},
t
\right)  
d\mathbf{q}
d\mathbf{q}\,^{\prime}
=
\sum\limits_{\nu}
G_{\rho\nu}\ast A_{\nu}.
\end{align*}
}

%\section{Discussion}\label{geo-singsoln}

\subsection{A geodesic Vlasov equation}
Importantly, geodesic motion for the $p-$moments is equivalent to
geodesic motion for the Euler-Poincar\'e equations on the
symplectomorphisms (EPSymp) given by the following Hamiltonian

\begin{equation}\label{epsymp}
H\left[  f\right]  =
\frac{1}{2}\iint f\left(q,p\right)  
\mathcal{G}\left(  q,p,q\,^{\prime},p\,^{\prime}\right)  
f\left(q\,^{\prime},p\,^{\prime}\right)  
dq\,dp\,dq\,^{\prime}dp\,^{\prime}
\end{equation}
The equivalence with EPSymp emerges when the function $\mathcal{G}$
is written as

\[
\mathcal{G}\left(  q,q\,^{\prime},p,p\,^{\prime}\right)  
=
\underset{n,m}{\sum}
\thinspace 
p^{n}\/G_{nm}\left(q,q\,^{\prime}\right) p\,^{\prime\, m}
\,.
\]
and the corresponding Vlasov equation reads as
\[
\frac{\partial f}{\partial t}\,+\,\Big\{f\,,\,\,\mathcal{G}*f\Big\}\,=\,0
\]
where $\{\cdot,\cdot\}$ denotes the canonical Poisson bracket.

Thus, whenever the metric $\mathcal{G}$ for EPSymp has a Taylor series,
its solutions may be expressed in terms of the geodesic motion for the
$p-$moments. More particularly the geodesic Vlasov equation presented here
is nonlocal in both position and momentum and
is equivalent to the vorticity equation in two-dimensions and for a particular choice of the metric. However this equation extends to more dimensions and to any kind of geodesic motion, no matters how the metric is expressed explicitly.

\rem{
Moreover the distribution function corresponding to the singular solutions
for the moments is a particular case of the {\bfi cold-plasma
approximation}, given by

\[
f(q,p,t)=\sum_j \rho_j(q,t)\,\delta(p-P_j(q,t))
\]
where in our case a summation is introduced and  $\rho$ is written as a Lagrangian particle-like
density:

\[
\rho_j(q,t)=\delta(q-Q_j(t))
\]
To check this is a solution for the geodesic motion of the generating function, one repeats exactly the same procedure as for the moments, in order to find the following Hamiltonian equations

\[
\frac {d Q_j}{d t}=\frac \partial {\partial \widetilde{P}_j}\,\,
\frac {\delta H}{\delta f} ( Q_j,\widetilde{P}_j ),
\qquad
\frac {d \widetilde{P}_j}{d t}=\frac \partial {\partial Q_j}\,\,
\frac {\delta H}{\delta f} ( Q_j,\widetilde{P}_j )
\]
where $\widetilde{P}_j=P_j\circ Q_j$ denotes the composition of the two functions
$P_j$ and $Q_j$. This procedure recovers single particle motion for density $\rho_j$ defined on a delta function.
%\comment{John, you seem to claim this is not a reduction, or vice versa,
%we could have reduced to this case long ago, since $P(q(t),t)$ is just
%$\tilde{P}(t)$ if $P(q(t),t)$ has no explicit time dependence. However, %I think we are allowing for explicit time dependence, too, don't you?
%\\Thanks -- Darryl }
\\
}

\subsection{Singular geodesic solutions}
We have now clarified the geometric meaning of the moment
equations and we can therefore characterize singular solutions, since the geodesic Vlasov equation (EPSymp) essentially describes advection in phase
space. Indeed, the geodesic Vlasov equation possesses the single
particle solution
\[
f(q,p,t)\,=\,\sum_j \,\delta(q-Q_j(t))\,\delta(p-P_j(t))
\]
which is a well known singular solution that is admitted whenever the phase-space
density is advected along a smooth Hamiltonian vector field.This happens, for example, in the Vlasov-Poisson system and in the general wakefield model. On the other hand, these singular solutions \emph{do not} appear in the Vlasov-Benney equation.

Equation (\ref{EPMS-eqn}) admits singular solutions of the form
\begin{eqnarray}
A_n({q},t)
&=&
\sum_{j=1}^N
P_j^n(t)\,
\delta\big({q}-{Q}_j(t)\big)
\label{sing-soln}
\end{eqnarray} 

In order to show this is a solution in one dimension, one checks that
these singular solutions satisfy a system of partial differential
equations in Hamiltonian form, 
%\begin{eqnarray*}
%\frac{d{Q}_j}{dt}
%=
%\frac{\partial H_N}{\partial P_j} 
%\,,&&
%\frac{\partial{P}_j}{\partial t}
%=-\,
%\frac{d{H}_N}{d{Q}_j} 
%\end{eqnarray*}
whose Hamiltonian couples all the moments
\begin{eqnarray*}
H_N
&=&
\frac{1}{2}\sum_{n,s=0}^\infty
\sum_{j,k=1}^N
P^s_j(t)\,P^n_k(t)\,
G_{ns}(Q_j(t),Q_k(t))
\end{eqnarray*}
Explicitly, one takes the pairing of the coadjoint equation
\[
\dot{A}_{m}
=
-\sum_{n,s}\mathrm{ad}_{G_{ns}\ast A_{s}}^{\ast}A_{m+n-1}
\]
with a sequence of smooth functions $\left\{  \varphi_{m}\left(  q\right) 
\right\}$
and finally obtains the equations for $Q_{j}$ and ${P}_{j}$ in
canonical form, 
\[
\frac{d Q_{j}}{d t}
=
\frac{\partial H_N}{\partial{P}_{j}},
\qquad
\frac{d{P}_{j}}{d t} 
=
-\,\frac{\partial H_N}{\partial Q_{j}}.
\]

These singular solutions of EPSymp are also
solutions of the Euler-Poincar\'e equations on the diffeomorphisms
(EPDiff). In the latter case, the single-particle solutions reduce to the pulson solutions for EPDiff \cite{CaHo1993}. Thus, the singular pulson solutions of the EPDiff equation arise naturally from the single-particle dynamics on phase-space. A similar result also holds in higher dimensions \cite{GiHoTr05}.

\paragraph{Further remarks on singular solutions.} Another kind of singular solution for the moments
may be obtained by considering the \emph{cold-plasma solution} of the Vlasov
equation
\[
f(q,p,t)=\sum_j\,\rho_j(q,t)\,\delta(p-P_j(q,t))
\] 
Indeed exchanging the variables $q\leftrightarrow p$ in the
single particle PDF leads to the following expression
\[
f(q,p,t)=\sum_j \psi_j(p,t)\delta(q-\lambda_j(p,t))
\]
which is always a solution of the Vlasov equation because of the symmetry
in $q$ and $p$. This leads to the following singular solutions for the moments:
\[
A_n(q,t)=\sum_j \int\!dp\, p^n\,\psi_{j}(p,t)\,\delta(q-\lambda_j(p,t))
\]
At this point, if one considers a Hamiltonian depending only on $A_1$ (i.e.
one considers the action of cotangent lifts of diffeomorphism), then it is
possible to drop the $p$-dependence in the $\lambda$'s and thereby recover to the
singular solutions previously found for eq. (\ref{conslaw}).

\rem{
\paragraph{Remark about higher dimensions}
The singular solutions (\ref{sing-soln}) with the
integrals over coordinates $a_j$ exist in higher
dimensions. The higher dimensional singular solutions satisfy a system of
canonical Hamiltonian integral-partial differential equations, instead of
ordinary differential equations.
} 

\rem{
\subsection{Exchanging variables in EPSymp}
One can show that exchanging the variables $q\leftrightarrow p$ in the
single particle PDF leads to another nontrivial singular
solution of EPSymp, which is different from those found previously. To
see this, let
$f$ be given by

\[
f(q,p,t)=\sum_j\delta(q-Q_j(p,t))\,\delta(p-P_j(t))
\]
At this stage nothing has changed with respect to the previous solution since
the generating function is symmetric with respect to q and p. However,
inserting this expression in the definition of the $m-$th moment yields

\[
A_m(q,t)=\sum_j\, P_j^m\,\delta(q-Q_j(P_j,t))
\] 
which is quite different from the solutions found previously. One
again obtains a canonical Hamiltonian structure for $P_j$ and $Q_j$.\\
This second expression is an alternative parametrisation of the cold-plasma
reduction above and it may be useful in situations where the composition
$Q_j\circ P_j$ is more convenient than $P_j\circ Q_j$.
}

\subsection{Examples of simplifying truncations and specializations.} 
The problem presented by the coadjoint motion equation  (\ref{A-dot}) for
geodesic evolution of $p-$moments under EPDiff needs further
simplification. One simplification would be to modify the (doubly)
infinite set of equations in (\ref{A-dot}) by truncating the Poisson bracket to a finite set. These moment dynamics may be truncated at any stage by modifying the Lie-algebra in the KMLP bracket to
vanish for weights $m+n-1$ greater than a chosen cut-off value.

For example, if we truncate the sums to $m,n=0,1,2$ only, then
equation (\ref{A-dot}) produces the coupled system of partial
differential equations,

\rem{
\begin{align*}
\frac{\partial A_{0}}{\partial t} &  =-\mathrm{ad}_{\beta_{1}}^{\ast}
A_{0}-\mathrm{ad}_{\beta_{2}}^{\ast}A_{1}\,,\\
\frac{\partial A_{1}}{\partial t} &  =-\mathrm{ad}_{\beta_{0}}^{\ast}
A_{0}-\mathrm{ad}_{\beta_{1}}^{\ast}A_{1}-\mathrm{ad}_{\beta_{2}}^{\ast}
A_{2}\,,\\
\frac{\partial A_{2}}{\partial t} &  =-\mathrm{ad}_{\beta_{0}}^{\ast}
A_{1}-\mathrm{ad}_{\beta_{1}}^{\ast}A_{2}\,.
\end{align*}
Expanding now the expression of the coadjoint operation
\[
\mathrm{ad}_{\beta_{h}}^{\ast}A_{k+h-1}=\left(  k+h\right)  A_{k+h-1}
\partial_{q}\beta_{h}+h\beta_{h}\partial_{q}A_{k+h-1}
\]
and relabeling
\[
\mathrm{ad}_{\beta_{h}}^{\ast}A_{k}=\left(  k+1\right)  A_{k}\partial_{q}
\beta_{h}+h\beta_{h}\partial_{q}A_{k}
\]
one calculates
}

\begin{align*}
\frac{\partial A_{0}}{\partial t} &  =-\partial_{q}\left(  A_{0}\beta_{1}\right)  -2A_{1}\partial_{q}\beta
_{2}-2\beta_{2}\partial_{q}A_{1}\\
\frac{\partial A_{1}}{\partial t} &  =-A_{0}\partial_{q}\beta_{0}-2A_{1}\partial_{q}\beta_{1}-\beta_{1}%
\partial_{q}A_{1}-3A_{2}\partial_{q}\beta_{2}-2\beta_{2}\partial_{q}A_{2}\\
\frac{\partial A_{2}}{\partial t} &  =-2A_{1}\partial_{q}\beta_{0}-3A_{2}\partial_{q}\beta_{1}-\beta_{1}%
\partial_{q}A_{2}%
\end{align*}

We specialize to the case that each velocity depends only on its
corresponding moment, so that $\beta_s=G*A_s$, $s=0,1,2$. If we further
specialize by setting $A_0$ and $A_2$ initially to zero, then these three
equations reduce to the single equation 
\begin{eqnarray*}
%\frac{\partial  A_0}{\partial t}
%&=&
%-\,{\partial_q}(A_0\beta_1)
%\,,\\
\frac{\partial  A_1}{\partial t}
&=&
-\,\beta_1\,{\partial_q}A_1
-\,2A_1\,{\partial_q}\beta_1
\,.%\label{A-dot2}
\end{eqnarray*}
Finally, if we assume that $G$ in the convolution $\beta_1=G*A_1$ is the
Green's function for the operator relation 
\[
A_1=(1-\alpha^2\partial_q^2)\beta_1
\]
for a constant lengthscale $\alpha$, then the evolution equation
for $A_1$ reduces to the integrable Camassa-Holm (CH) equation
\cite{CaHo1993} in the absence of linear dispersion. This is the one-dimensional
EPDiff equation, which has singular (peakon) solutions. Thus, after these
various specializations of the EPDiff $p-$moment equations, one finds the
integrable CH peakon equation as a further specialization of the coadjoint moment
dynamics of equation (\ref{A-dot}).

That such a drastic restriction of the $p-$moment system still leads to
such an interesting special case bodes well for future investigations of
the EPSymp $p-$moment equations. Further specializations and truncations
of these equations will be explored elsewhere. Before closing, we mention
one or two other open questions about the solution behavior of the
$p-$moments of EPSymp.

\section{Open questions for future work}
\paragraph{Emergence of singular solutions.} Several open questions remain for future work. The first of these is whether the singular solutions found here will emerge
spontaneously in EPSymp dynamics from a smooth initial Vlasov PDF. This
spontaneous emergence of the  singular solutions does occur for EPDiff.
\rem{
Namely, one sees the singular solutions of EPDiff emerging from {\it any}
confined initial distribution of the dual variable. (The dual variable is
fluid velocity in the case of EPDiff).
}
In fact, integrability of EPDiff in
one dimension by the inverse scattering transform shows that {\it only}
the singular solutions (peakons) are allowed to emerge from any confined
initial distribution in that case \cite{CaHo1993} (this also happens in higher
dimensions as it is shown by numerical simulations).
\rem{
In higher dimensions, numerical
simulations of EPDiff show that again only the singular solutions emerge
from confined initial distributions. 
}
In contrast, the point vortex
solutions of Euler's fluid equations (which are isomorphic to the cold
plasma singular solutions of the Vlasov-Poisson equation) while
comprising an invariant manifold of singular solutions, do not
spontaneously emerge from smooth initial conditions. Nonetheless, something quite analogous to the singular solutions
is seen experimentally for cold plasma in a Malmberg-Penning trap \cite{DuON1999}.
Therefore, one may ask which outcome will prevail for the singular
solutions of EPSymp. Will they emerge from a confined smooth initial
distribution, or will they only exist as an invariant manifold for
special initial conditions? One might argue that in two dimensions, the EPSymp
equation encompasses the equation of vorticity and thus spontaneous
emergence of point vortices should not occur. However it is possible that
the choice of the metric plays an important role in this matter. Of course, the interactions of these singular
solutions for various metrics and the properties of their collective
dynamics is a question for future work.

\paragraph{Possible connections with the Bloch-Iserles system} The EPSymp equation is surprisingly similar in construction to another important integrable geodesic equation on the linear Hamiltonian vector
fields (Hamiltonian matrices), which has recently been proposed \cite{BlIs}.
This is a finite dimensional equation whose dynamical variables are symmetric
matrices. Now it has been shown that this system may be written as the geodesic equation on the group of the linear canonical transformations Sp$( \mathbb{R} ;2n )$
\cite{BlIsMaRa05}.
This association to canonical transformations raises the question wether it is possible to establish connections
with the geodesic Vlasov equation that was introduced here

\subsection*{Acknowledgements} 
%\todo{To be revised}
CT is grateful to Ugo Amaldi and Riccardo Zennaro for helpful discussions on the wakefield model for accelerator beams. The work of DDH was partially supported by US DOE, Office of Science, Applied Mathematics Research program.

%%%%%%%%%%%%%%%%%%%%%
\bibliographystyle{unsrt}

\begin{thebibliography}{999}
\itemsep -1mm


\bibitem{Be1973}
Benney, D. J. \textit{Properties of long nonlinear waves}.
Stud. App. Math. 52 (1973) 45. 

\bibitem{BlBrGlRaRy03}
Blaskiewicz, M.; Brown, K.; Glenn, J. W.; Raka, E.; Ryan, J. {\it Spill structure in intense beams.} Proceedings of the Particle Accelerator Conference Vol.
4 (2003), 2595--2597 

\bibitem{BlIs}
Bloch, A. M.; Iserles, A. {\it On an isospectral Lie-Poisson system and its Lie algebra.} Found. Comput. Math. 6 (2006), no. 1, 121--144

\bibitem{BlIsMaRa05}
Bloch, A. M.; Iserles, A.; Marsden, J.E.; Ratiu T. S.
\textit{A class of integrable geodesic flows on the symplectic group and the symmetric matrices.} arXiv.org:math-ph/0512093 (2005).

\bibitem{CaHo1993}
Camassa, R.; Holm, D. D. \textit{An integrable shallow water equation
with peaked solitons.} Phys. Rev. Lett. 71 (1993), no. 11, 1661--1664.

\bibitem{CaMaPu2002}
Caprino, S.; Marchioro, C.; Pulvirenti, M.
{\it On the two-dimensional Vlasov-Helmholtz equation with infinite mass.} Comm. Partial Differential Equations 27 (2002), no. 3-4, 791--808

\bibitem{CeHoHoMa1998}
Cendra, H.; Holm, D. D.; Hoyle, M. J. W.; Marsden, J. E. \textit{The
Maxwell-Vlasov equations in Euler-Poincar\'e form.} J. Math. Phys. 39
(1998), no. 6, 3138--3157.

\bibitem{CeMaPeRa}
Cendra, H.; Marsden, J. E.; Pekarsky, S.; Ratiu, T. S.
\textit{Variational principles for Lie-Poisson and Hamilton-Poincar\'e
equations.} Mosc. Math. J. 3 (2003), no. 3, 833--867.

\bibitem{CoDaHoMa04}
Cousineau, S.; Danilov, V.; Holmes, J.; Macek, R. {\it Space-charge-sustained
microbunch structure in the Los Alamos Proton Storage Ring.} Phys. Rev. ST Accel. Beams 7 (2004), no. 9

\bibitem{Dragt} 
Dragt, A. J.; Neri, F.; Rangarajan, G.; Douglas, D. R.; Healy, L. M.;
Ryne, R. D. \textit{Lie algebraic treatment of linear and nonlinear beam
dynamics.} Ann. Rev. Nucl. Part. Sci. 38 (1990), no. 38, 455--496

\bibitem{DuON1999} 
Dubin, D. H. E.; O'Neil, T. M.
\textit{Trapped nonneutral plasmas, liquids, and crystals (the thermal
equilibrium states)}.
Rev. Mod. Phys., 71 (1990), no. 1, 87-172.

\bibitem{Gi1981}
Gibbons, J. \textit{Collisionless Boltzmann equations and integrable
moment equations.} Phys. D 3 (1981), no. 3, 503--511.

\bibitem{GiHoKu1982}
Gibbons, J.; Holm, D. D.; Kupershmidt, B. A. \textit{Gauge-invariant
Poisson brackets for chromohydrodynamics.} Phys. Lett. A 90 (1982), no. 6,
281--283.

\bibitem{GiHoKu1983} 
Gibbons, J.; Holm, D. D.; Kupershmidt, B. A. \textit{The Hamiltonian
structure of classical chromohydrodynamics.} Phys. D 6 (1982/83), no. 2,
179--194.

\bibitem{GiHoTr05}
Gibbons, J.; Holm, D. D.; Tronci, C. {\it Singular solutions for geodesic flows of Vlasov moments}, Proceedings of the MSRI workshop \textit{Probability, geometry and integrable systems}, in celebration of Henry McKean's $75^\text{th}$ birthday, Cambridge University Press, Cambridge, 2007 (in press, also at arXiv.org:nlin/0603060)

\bibitem{HoMa2004} 
Holm, D. D.; Marsden, J. E. 
\textit{Momentum maps and measure valued solutions (peakons, filaments,
and sheets) of the Euler-Poincar«e equations
for the diffeomorphism group.}
In {\it The Breadth of Symplectic and Poisson Geometry, A Festshrift for
Alan Weinstein}, 203-235, { Progr. Math.}, {232}, J.E. Marsden
and T.S. Ratiu, Editors, Birkh\"auser Boston, Boston, MA, 2004. 

\bibitem{HoMaRa}
Holm, D. D.; Marsden, J. E.; Ratiu, T. S. \textit{The Euler-Poincar\'e
equations and semidirect products with applications to continuum
theories.} Adv. Math. 137 (1998), no. 1, 1--81.

\bibitem{HoLySc1990}
Holm, D. D.; Lysenko, W. P.; Scovel, J. C. \textit{Moment invariants for
the Vlasov equation.} J. Math. Phys. 31 (1990), no. 7, 1610--1615.

\bibitem{HoRaTrYo2004}
Holm, D. D.; Ratnanather, T. J.; Trouvé, A.; Younes, L. {\it Soliton dynamics in computational anatomy}, Neuroimage 23 (2004) S170--S178

\bibitem{HoSt03}
Holm, D. D.; Staley, M. F. \textit{Wave structure and nonlinear balances in a family of evolutionary PDEs.} SIAM J. Appl. Dyn. Syst. 2 (2003), no. 3

\bibitem{KoHaLi2001}
Koscielniak, S.; Hancock, S.; Lindroos, M. {\it Longitudinal holes in debunched particle beams in storage rings, perpetuated by space-charge forces.} Phys. Rev. ST Accel. Beams 4 (2001), no. 4

\bibitem{KuMa}
Kupershmidt, B. A.; Manin, Ju. I. \textit{Long wave equations with a
free surface. II. The Hamiltonian structure and the higher equations.}
Funktsional. Anal. i Prilozhen. 12 (1978), no. 1, 25--37

\bibitem{LeMa}
Lebedev, D.R.; Manin, Ju. I. \textit{Conservation laws and Lax
representation of Benney's long wave equations.} Phys. Lett. A 74 (1979),
no. 3-4, 154--156.

\bibitem{MaRa99}
Marsden, J. E.; Ratiu, T. S. {\it Introduction to mechanics and symmetry. A basic exposition of classical mechanical systems.} Second edition. Texts in Applied Mathematics, 17. Springer-Verlag, New York, 1999.

\bibitem{MaWe}
Marsden, J. E.; Weinstein, A. \textit{The Hamiltonian structure of the
Maxwell-Vlasov equations.} Phys. D 4 (1981/82), no. 3, 394--406.

\bibitem{MaWeRaScSp}
Marsden, J. E.; Weinstein, A.; Ratiu, T.; Schmid, R.; Spencer, R. G.
\textit{Hamiltonian systems with symmetry, coadjoint orbits and plasma
physics.} Proceedings of the IUTAM-ISIMM symposium on modern developments
in analytical mechanics, Vol. I (Torino, 1982). Atti Accad. Sci. Torino
Cl. Sci. Fis. Mat. Natur. 117 (1983), suppl. 1, 289--340.

\bibitem{MoBaJaLeNgShTa}
Moore, R.; Balbekov, V.; Jansson, A.; Lebedev, V.; Ng, K.Y.; Shiltsev, V.; Tan, C.Y. {\it Longitudinal bunch dynamics in the Tevatron.} Proceedings
of the Particle Accelerator
Conference 
Vol. 3 (2003), 1751--1753 

\bibitem{QiTa2004}
Qin, H.; Tang, W.M. {\it Pullback transformations in gyrokinetic theory.}
Phys. of Plasmas 11 (2004), no. 3, 1052--1063

\bibitem{ScWe}
Scovel, C.; Weinstein, A. \textit{Finite-dimensional Lie-Poisson
approximations to Vlasov-Poisson equations.} Comm. Pure Appl. Math. 47
(1994), no. 5, 683--709.

\bibitem{WeMo}
Weinstein, A.; Morrison, P. J. \textit{Comments on:
``The Maxwell-Vlasov equations as a continuous Hamiltonian system''.}
Phys. Lett. A 80 (1980), no. 5-6, 383--386.

\rem{
\bibitem{ArKe98}
Arnold, V. I.; Khesin, B. A. \textit{Topological methods 
in hydrodynamics.} 
Applied Mathematical Sciences, 125, Springer-Verlag, New York, 1998.

\bibitem{ElRa91}
Eliashberg, Y.; Ratiu, T. \textit{The diameter of the symplectomorphism group
is infinite.} Invent. Math. 103 (1991), no. 2, 327--340.
}

\bibitem{ScFe2000}
Schamel, H.; Fedele, R. {\it Kinetic theory of solitary waves on coasting beams in synchrotrons.} Phys. Plasmas 7 (2000), no. 8, 3421--3430 

\rem{
\bibitem{Shn85}
Shnirelman, A. I. \textit{The geometry of the group of diffeomorphisms and the dynamics of an ideal incompressible fluid.} Mat. Sb. 128 (1985),
no. 1, 82--109, 144. 
}

\bibitem{Venturini}
Venturini, M. {\it Stability analysis of longitudinal beam dynamics using noncanonical Hamiltonian methods and energy principles.} Phys. Rev. ST Accel. Beams 5 (2002), no. 5


\end{thebibliography}
%\todo{Check that each reference is properly cited}

%%%%%%%%%%%%%%%%%%%%%

\end{document}